\documentclass[10pt, conference, letterpaper]{IEEEtran}

\usepackage{cite}
\usepackage{amsmath,amssymb,amsfonts}
\usepackage{algorithmic}
\usepackage{graphicx}
\usepackage{overpic}
\usepackage{textcomp}
\usepackage{xcolor}
\def\BibTeX{{\rm B\kern-.05em{\sc i\kern-.025em b}\kern-.08em
    T\kern-.1667em\lower.7ex\hbox{E}\kern-.125emX}}

\usepackage{orcidlink}
\usepackage[nolist,nohyperlinks]{acronym}
\usepackage[detect-weight=true, detect-family=true]{siunitx}
\DeclareSIUnit{\belmilliwatt}{Bm}
\DeclareSIUnit{\bel}{B}
\DeclareSIUnit{\bitpersecond}{bps}
\DeclareSIUnit{\samplepersecond}{Sps}
\DeclareSIUnit{\nothing}{\relax}

\usepackage{placeins}
\usepackage[inkscapelatex=false]{svg}
\usepackage{multirow}
\usepackage{enumerate}
\usepackage[inline]{enumitem}
\usepackage{hyperref}
\usepackage[nameinlink, capitalise]{cleveref}
\usepackage{comment}
\makeatletter
\newcommand*{\org@overidelabel}{}
\let\org@overridelabel\@verridelabel
\@ifpackagelater{acronym}{2015/03/21}{
  \renewcommand*{\@verridelabel}[1]{%
    \@bsphack
    \protected@write\@auxout{}{\string\AC@undonewlabel{#1@cref}}%
    \org@overridelabel{#1}%
    \@esphack
  }%
}{
  \renewcommand*{\@verridelabel}[1]{%
    \@bsphack
    \protected@write\@auxout{}{\string\undonewlabel{#1@cref}}%
    \org@overridelabel{#1}%
    \@esphack
  }%
}

\newcommand{\linebreakand}{%
  \end{@IEEEauthorhalign}
  \hfill\mbox{}\par
  \mbox{}\hfill\begin{@IEEEauthorhalign}
}
\makeatother
\begin{acronym}
    \acro{IoT}{Internet of Things}
    \acro{MiniRocket}[\textsc{MiniRocket}]{\textbf{MINI}mally \textbf{R}and\textbf{O}m \textbf{C}onvolutional \textbf{KE}rnel \textbf{T}ransform}
    \acro{PPV}{proportion of positive values}
    \acro{MCU}{microcontroller unit}
    \acro{RF}{radio frequency}
    \acro{NFC}{near-field communication}
    \acro{I2C}{Inter-Integrated Circuit}
    \acro{SoC}{system-on-chip}
    \acro{FFT}{fast fourier transform}
    \acro{TEG}{thermoelectric generator}
    \acro{MEMS}{microelectromechanical system}
    \acro{BLE}[\textsc{BLE}]{\textsc{Bluetooth Low Energy}}
    \acro{ML}{machine learning}
    \acro{HAR}{human activity recognition}
    \acro{CNN}{convolutional neural network}
    \acro{DT}{decision tree}
    \acro{GBC}{gradient boosting}
    \acro{SVM}{support vector machine}
    \acro{kNN}{k nearest neighbours}
    \acro{PRNG}{pseudorandom number generator}
    \acro{FPU}{floating point unit}
    \acro{UTC}{coordinated universal time}
    \acro{FIR}{finite impulse response}
    \acro{PC}{personal computer}
    \acro{IMU}{inertial measurement unit}
\end{acronym}

\usepackage{eso-pic}
\usepackage{url}
\AddToShipoutPictureBG*{
  \AtPageUpperLeft{%
    \put(0,-40){\raisebox{15pt}{\makebox[\paperwidth]{\begin{minipage}{21cm}\centering
      \textcolor{gray}{This article has been accepted for publication in the proceedings of the 2023 IEEE 9th World Forum on Internet of Things (WFIoT).} 
    \end{minipage}}}}%
  }
  \AtPageLowerLeft{%
    \raisebox{25pt}{\makebox[\paperwidth]{\begin{minipage}{21cm}\centering
      \textcolor{gray}{ \copyright 2023 IEEE.  Personal use of this material is permitted.  Permission from IEEE must be obtained for all other uses, in any current or future media, including reprinting/republishing this material for advertising or promotional purposes, creating new collective works, for resale or redistribution to servers or lists, or reuse of any copyrighted component of this work in other works.
      }
    \end{minipage}}}%
  }
}

\begin{document}
\title{Optimizing \acs{IoT}-Based Asset and Utilization Tracking: Efficient Activity Classification with \acs{MiniRocket} on Resource-Constrained Devices}

\author{
    \IEEEauthorblockN{
        Marco Giordano\IEEEauthorrefmark{1}\orcidlink{0000-0003-1106-3472}, \textit{Graduate Student Member, IEEE}, 
        Silvano Cortesi\IEEEauthorrefmark{1}\orcidlink{0000-0002-2642-0797}, \textit{Graduate Student Member, IEEE}, \\
        Michele Crabolu\IEEEauthorrefmark{2}\orcidlink{0000-0003-4308-2432},
        Lavinia Pedrollo\IEEEauthorrefmark{2},
        Giovanni Bellusci\IEEEauthorrefmark{2}\orcidlink{0000-0002-5514-2503},
        Tommaso Bendinelli\IEEEauthorrefmark{3},
        \\Engin Türetken\IEEEauthorrefmark{3}, 
        Andrea Dunbar\IEEEauthorrefmark{3} and 
        Michele Magno\IEEEauthorrefmark{1}\orcidlink{0000-0003-0368-8923}, \textit{Senior Member, IEEE}}
    \IEEEauthorblockA{\IEEEauthorrefmark{1}\textit{Dept. of Information Technology and Electrical Engineering, ETH Zürich, Zürich, Switzerland,} name.surname@pbl.ee.ethz.ch}
    \IEEEauthorblockA{\IEEEauthorrefmark{2}\textit{Hilti Corporation, Schaan, Liechtenstein,}
    name.surname@hilti.com}
    \IEEEauthorblockA{\IEEEauthorrefmark{3}\textit{CSEM SA, Neuchâtel, Switzerland,}
    name.surname@csem.ch}
}

\maketitle
\begin{abstract}

This paper introduces an effective solution for retrofitting construction power tools with low-power \ac{IoT} to enable accurate activity classification. We address the challenge of distinguishing between when a power tool is being moved and when it is actually being used. To achieve classification accuracy and power consumption preservation a newly released algorithm called \ac{MiniRocket} was employed. Known for its accuracy, scalability, and fast training for time-series classification, in this paper, it is proposed as a TinyML algorithm for inference on resource-constrained \ac{IoT} devices.
The paper demonstrates the portability and performance of \ac{MiniRocket} on a resource-constrained, ultra-low power sensor node for floating-point and fixed-point arithmetic, matching up to 1\% of the floating-point accuracy.
The hyperparameters of the algorithm have been optimized for the task at hand to find a Pareto point that balances memory usage, accuracy and energy consumption.
For the classification problem, we rely on an accelerometer as the sole sensor source, and \ac{BLE} for data transmission.
Extensive real-world construction data, using 16 different power tools, were collected, labeled, and used to validate the algorithm's performance directly embedded in the \ac{IoT} device.
Experimental results demonstrate that the proposed solution achieves an accuracy of \(96.9\%\) in distinguishing between real usage status and other motion statuses while consuming only \qty{7}{\kilo\byte} of flash and \qty{3}{\kilo\byte} of RAM. The final application exhibits an average current consumption of less than \qty{15}{\micro\watt} for the whole system, resulting in battery life performance ranging from 3 to 9 years depending on the battery capacity (\(\qtyrange[range-phrase = -, range-units = single]{250}{500}{\milli\ampere{}\hour}\)) and the number of power tool usage hours (\(\qtyrange[range-phrase = -, range-units = single]{100}{1500}{\hour}\)).
\end{abstract}


\acresetall
\section{Introduction}

In the fast-evolving world of the \ac{IoT}~\cite{sharma23_data_conver_desig_space_explor_iot_applic, m.15_evolut_iot, cano18_evolut_iot}, a new generation of intelligent sensor nodes is transforming how we understand and interact with our environment~\cite{sharma18_histor_presen_futur_iot}. These nodes exploit advanced algorithms to extract information from the data collected directly close to the sensors, instead of merely collecting and transmitting data to a centralized server~\cite{cao2020overview}. This significant enhancement in processing the information at the edge brings a new level of efficiency, responsiveness, and adaptability to a broad range of applications, from smart homes~\cite{babangida22_inter_thing_iot_based_activ} to industrial automation~\cite{khan20_indus_inter_thing} and beyond, pointing to a promising future for \ac{IoT} technology~\cite{brincat2019internet,salam2019internet}.
Among other applications of \ac{IoT}, one promising in the industrial domain is asset- and utilization-tracking, aiming to optimize the asset park, health, and usage of physical assets such as machinery, vehicles, or power tools~\cite{saxena2020iot}.
\par
Optimizing assets, such as power tools in the construction sector, is crucial for achieving higher productivity and sustainability standards~\cite{saxena2020iot}. Retrieving information on their utilization and health becomes therefore essential. Activity classification can play a crucial role for achieving such objectives. The advancements in \ac{ML} have enabled the development of solutions that offer accurate detection using a variety of algorithms that operate on sensor data~\cite{ha2020machine,wang2020fann}.
\par
In order to run \ac{ML} models on the node, we need to collect and process data on the fly, requiring an advanced hardware/software co-design. Incorporating sensor nodes within the tool presents several challenges: different companies would need to develop their own systems; space constraints within the tools often limit the options; strict certifications must be met; and retrofitting older tools becomes difficult or even impossible. Furthermore, adding electronics inside the tool leads to significant costs due to variations in tool design, functionality, and space constraints, depending on the specific tool~\cite{kanan18_iot_based_auton_system_worker}.
Alternatively, utilizing an external device for monitoring purposes can be a better alternative. However, this approach brings its own set of challenges. Firstly, the external device relies on its own power supply, necessitating a long battery life for usability and cost-effectiveness. This energy boundary limits the computational resources of the processing units. Secondly, there is no direct access to the tool's information, requiring data to be collected, processed, and transmitted solely through external sensors~\cite{ha2020machine}. This limits the possible physical phenomena that can be sensed, making the activity classification task harder. Additionally, the cost of components and manufacturing has also to be considered, adding another level of complexity to the design.
\par
In light of these challenges, our research aims to contribute to the field by proposing a novel approach that preserves scalability and addresses the limitations of existing \ac{ML} approaches to run on ultra-low power \acp{MCU} with few \unit{\kilo\byte} of memory. We target a middle ground of model expressiveness and computational complexity, aiming for more complex models than naive threshold-based classifiers, without having to deal with the hefty requirements of neural networks. We propose a solution that leverages a newly released algorithm called \acf{MiniRocket}. \ac{MiniRocket} is a multi-class time series classifier, recently introduced by Dempster et al.~\cite{dempster21_minir}. \ac{MiniRocket} has been introduced as an accurate, fast, and scalable training method for time-series data, requiring remarkably low computational resources to train. It has been shown to achieve state-of-the-art accuracy on a large number of time series datasets from the UCR archive~\cite{UCRArchive}. We propose to utilize its low computational requirements as a TinyML algorithm for resource-constrained \ac{IoT} devices. Our goal is to demonstrate that the algorithm would prove lightweight also for inference and is therefore very well fitting resource-constrained \ac{IoT} devices with low-power processing cores,  such as \textsc{ARM Cortex-M4F} cores, nowadays integrated into almost all \ac{BLE} \acp{SoC}~\cite{cerutti2020sound}, which has been proved to run inference at the edge\cite{9870017}. By utilizing \ac{MiniRocket}'s low computational requirements~\cite{dempster21_minir}, our approach enables accurate activity detection by placing a sensor outside of the power tools, enabling a simple and cheap retrofitting on already manufactured tools.
Moreover, using an algorithm that learns features removes the need for human intervention and adaption to different tasks and/or different data, making an algorithm such as \ac{MiniRocket} better at generalization and future-proofing.
To the best of our knowledge, this is the first work to have ported the \ac{MiniRocket} algorithm to \textsc{C}, providing both floating point and fixed point implementations, and run it on an \ac{MCU}.
\par
With the goal of bringing intelligence in a compact and ultra-low power tag, in this work, the {MiniRocket} algorithm has been efficiently ported on a low-power \ac{MCU}. A sensor node previously designed in~\cite{giordano21_smart,giordano22_desig_perfor_evaluat_ultral_power} has been used for an extensive evaluation of the embedded performance and power consumption, demonstrating the feasibility of advanced activity recognition in low-power processors. We decided to base our evaluation on accelerometer data only, as they are available at low cost and require very little power (e.g. \qty{5}{\micro\ampere} at \qty{100}{\hertz} sampling rate in the case of the \textsc{IIS2DLPCT} used later). 
In particular, in order to track an asset's working life, so that its correct use, productivity, wear, and associated maintenance can be ensured in a timely fashion, we propose to estimate the exact runtime of the tool~\cite{zonta20_predic_maint_indus}.
\par
The main contributions of this work are as follows:
\begin{enumerate}[label=(\roman*),font=\itshape] 
    \item Implementation of \ac{MiniRocket} in \textsc{C} in both floating point and fixed point mathematics, targeting low-power microcontrollers and deployment on a resource-constrained ultra-low power sensor node;
    \item Accurate evaluation of the fixed-point implementation of the \ac{MiniRocket} algorithm on a resource-constrained \ac{IoT} device - profiling especially memory and power.
    \item Extensive data collection and labeling of accelerometer data, recorded on 16 different power tools from different manufacturers performing 12 different activities. Training and validation of \ac{MiniRocket} on a classification problem. Analysis of optimal algorithm settings, and sensor requirements to optimize power consumption, memory, and computational requirements;
\end{enumerate}

The remainder of the paper is structured as follows: \cref{sec:relatedworks} presents the recent literature in asset- and utilization-tracking with a focus on activity detection and runtime estimation; \cref{sec:materials} introduces the experimental setup, the implemented algorithm, and its optimizations; \cref{sec:results} shows the results evaluated in a real-world scenario; Finally, \cref{sec:conclusion} concludes the paper.
\section{Related Works}\label{sec:relatedworks}

Since the widespread availability of \ac{IoT} in the industrial sector, monitoring devices using accelerometers have become more common~\cite{magno2019self} . Most of these use the nodes only to acquire data for later post-processing~\cite{caviezel2017design}, but not for complex data analysis in real-time, which requires advanced algorithms~\cite{labelle00_mater_class_by_drill, kanan18_iot_based_auton_system_worker}.
Previous work has shown that asset tracking is possible, especially for fault diagnosis. Magno et al.~\cite{magno19_smart} presented in their work a blade-wear detection task with on-node data processing and fault diagnosis. Data was recorded by an accelerometer, processed on a \textsc{Texas Instruments MSP430} by calculating the mean absolute value, comparing it with a threshold, and then transmitted it to a computer via \textsc{ZigBee}.
\par
In~\cite{gondo20_accel_based_activ_recog_worker, joshua11_accel_based_activ_recog_const, calvetti20_worker}, activity recognition is performed at the construction site, although not on the tool itself but on the worker. Each construction worker carries a sensor node with him, which is equipped with an accelerometer. The goal is to determine when a worker is active or inactive, in order to measure and control safety, productivity, and quality in construction sites. However, the algorithm evaluation is carried out only on \ac{PC}, without providing an embedded implementation.
\par
Using a neural network approach, in~\cite{labelle00_mater_class_by_drill} the authors classify different types of rocks while drilling. To do so, the authors evaluated signals from five different sensors, while in our work we try to rely only on accelerometer readings, in order to decrease cost and power consumption. Again, the authors do not target any embedded implementation. 
In~\cite{kanan18_iot_based_auton_system_worker}, the sensors needed to be integrated into the drill body, and the signals were collected and analyzed by external equipment, therefore the algorithm was not subject to any low-power requirements. And, in fact, they used hefty dense layers that would be too big for a resource-constrained device.
In~\cite{doerr19_recog_produc_applic_based_integ}, Dörr et al. tried to analyze a grinder and a cordless screwdriver using an \ac{IMU} and magnetometers. Gyroscopes and magnetometers are more expensive and not as power-efficient as accelerometers. They proposed a series of algorithms based first on a broad feature selection followed by various \ac{ML} approaches such as \ac{DT}, \ac{GBC}, \ac{kNN} and \ac{SVM} in order to perform an activity detection. However, their \ac{ML} algorithm was developed and run on a \ac{PC}, not targeting embedded devices.
\par
An on-device detection of the activity of power tools has already been attempted in previous work by the same authors~\cite{giordano21_smart, giordano22_desig_perfor_evaluat_ultral_power}. In those works, the detection has first been done by thresholding the \ac{FFT} extracted from accelerometer data, and later also by employing deep learning on the edge with a \ac{CNN}-based approach. Different activities, such as walking, and drilling through wood, or steel, could be detected and distinguished with an accuracy of \(90.6\%\). However, the algorithm analysis was limited to a single tool, and a neural network was chosen as the classifier algorithm. In this study, we collected a much bigger dataset, that can prove better the performance of the algorithm. Additionally, we employ a novel classification algorithm, \ac{MiniRocket}~\cite{dempster21_minir}, validate it, and implement it with fixed-point arithmetic, with the aim to further optimize the memory footprint and low-power operation.
\par
The existing literature explores various approaches to activity detection using accelerometers and other sensors in different contexts. However, the proposed solution in this paper stands out by leveraging the efficient and accurate \ac{MiniRocket}~\cite{dempster21_minir} algorithm, which has been ported in \textsc{C} with fixed-point support and evaluated extensively.  Moreover, \ac{MiniRocket} can generalize to any time series for both classification and regression tasks, paving the way to future studies and more complex \ac{ML} tasks.
Additionally, in this study, we contribute to the field by collecting and utilizing a real-world dataset for evaluation. Unlike previous works that often rely on lab-generated or simulated datasets, our dataset consists of extensive real-world construction data collected using 16 different power tools. This approach ensures a more realistic and accurate evaluation of the proposed solution. Ultimately, the algorithm has been implemented on an ultra-low power embedded device, where data are collected and processed in real-time.
\section{Materials and Methods}\label{sec:materials}
In this section, data collection setup, hardware, and software components of the \ac{IoT} sensor node will be introduced, along with the \ac{MiniRocket} algorithm and its efficient implementation.

\subsection{Data collection}
To ensure a diverse and encompassing data set, 16 different tools, categorized as jigsaw, circular saw, gas-powered saw, combi hammer, diamond cording, and breakers, were selected from six different brands and six different families. This diverse selection helps in capturing a wider range of variables that could influence the data and hence the algorithm's performance, ultimately assessing the ability to generalize, proving our important point of tool retrofitting.
\begin{figure}[!t]
  \centering
  \includegraphics[width=0.45\textwidth,trim={0 0.3cm 0 1.5cm},clip]{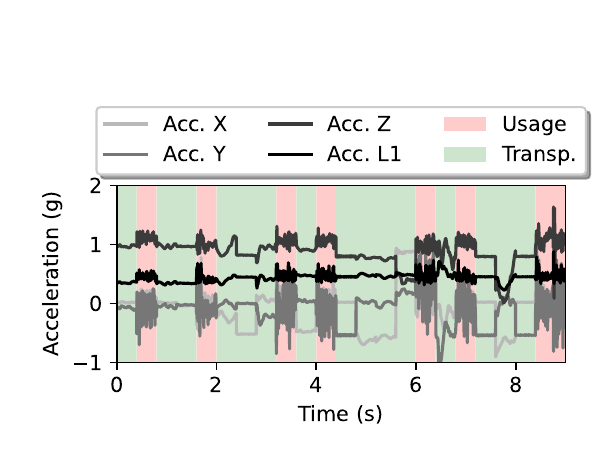}
  \caption{Plot showing a collected sample together with its labeling.}
  \label{fig:data_collected}
  \vspace{-0.5cm}
\end{figure}
\begin{description}
    \item[Hardware:] Data collection was conducted using the \textsc{ENDAQ S2-D8D16}, a standalone sensor data logger, attached to each power tool. The sensor was configured to output accelerometer data at \qty{3.2}{\kilo\hertz}. A digital camera was used to record every working activity. The videos were subsequently used to precisely label each frame along with the recorded sensor data. 
    \item[Protocol:]  Seven distinct working activities, ranging from wood sawing to concrete cutting, and five different modes of transportation, such as car driving and walking, were executed for data collection. The objective here was to record data in mixed classes and transition phases, simulating real-time and real working conditions. For each of these activities, active working (\textit{Usage} class) and any other non-active working phase present during the construction activity (here called \textit{Transportation} class) were blended in the same recording and then labeled accordingly sample by sample. A representation of some sample data collected can be seen in Figure \cref{fig:data_collected}.
    \item[Labelling:] A key part of the process was the alignment of sensor data with the video. At the beginning of each collection, the sensor was tapped five times, and the  fifth tap's \ac{UTC} timestamp was identified in both the accelerometer signal and the camera, and both sensor signals and video were trimmed to begin from this moment. This meticulous synchronization allowed for a precise match between video frames and sensor signals. Data labeling was facilitated by a script that plays the video frame by frame, allowing the user to label each frame using hotkeys from a keyboard. This step resulted in every timestamp being labeled with a class. Data pre-processing and synchronization involved resampling data to achieve an exact sample rate, given the discovered jitter of approximately \(1\%\) in the sensor's sampling frequency. Before resampling the time series from original \qty{3.2}{\kilo\hertz} to lower rates, a low-pass \ac{FIR} anti-aliasing filter was applied. This comprehensive methodology resulted in a high-quality dataset.
    \item[Windowing and splitting:] Data was split into train and test sets and windowed accordingly. In order to test the generalization capability of the algorithm, data was clustered based on the power tool brand as follows: training data contained only data from a single brand, i.e., \textsc{Hilti}; the remaining data collected with all other tool brands were preserved for the test set. The purpose of this split is to test the \ac{ML} algorithm in a very challenging condition, validating its ability to generalize towards different tools and companies. While performing the hyperparameter search, optimizing the sampling frequency, the number of features, and window length, data were windowed. \qty{20}{\kilo\nothing} and \qty{2}{\kilo\nothing} windows were randomly selected for the training and validation sets respectively. To ensure balance, half of the windows were chosen from the \textit{Transportation} class, while the remaining half were chosen from the \textit{Usage} class. Windows with mixed labels were discarded.
\end{description}
\begin{figure*}[!htpb]
    \centering
    \begin{overpic}[width=0.28\linewidth]{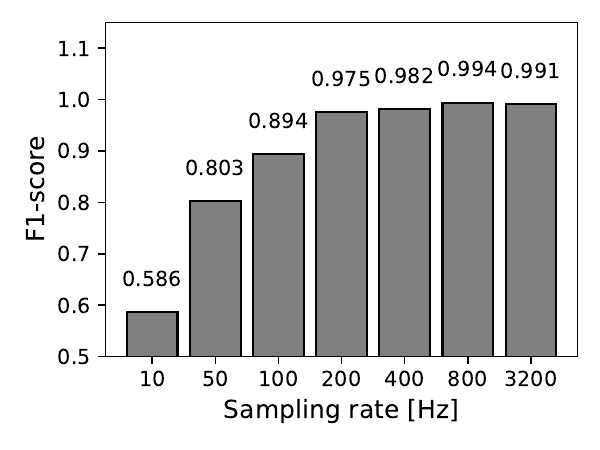}
        \put(5,1){(a)}
    \end{overpic}
    \begin{overpic}[width=0.28\linewidth]{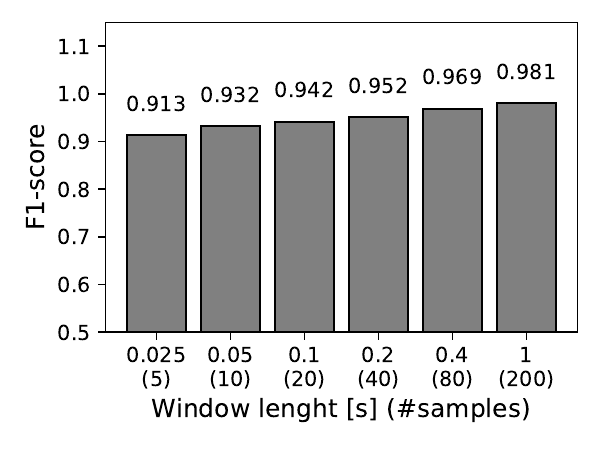}
        \put(5,1){(b)}
    \end{overpic}
    \begin{overpic}[width=0.42\linewidth]{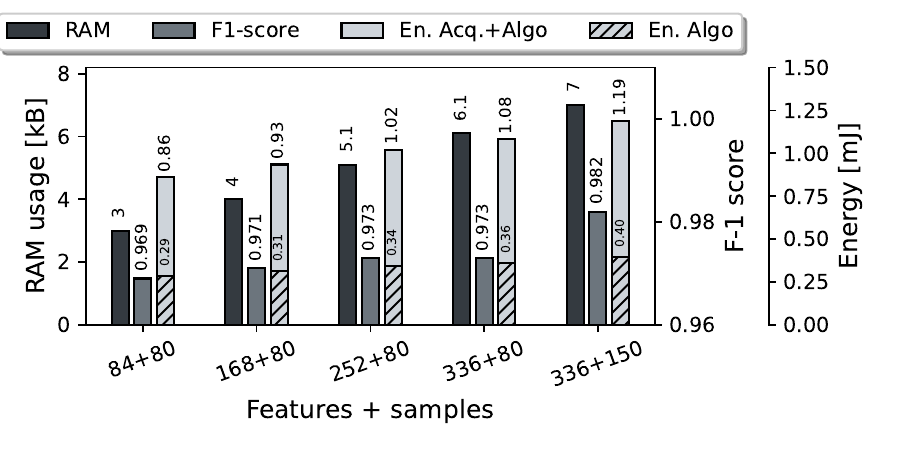}
        \put(5,1){(c)}
    \end{overpic}
    \caption{Plots showing the algorithm's performance, varying the (a) sampling rate and (b) window length. Plot (c) shows the memory consumption and F1-score varying features and samples.}
    \label{fig:algo_tuning}
    \vspace{-0.4cm}
\end{figure*}

\subsection{Algorithm}
The \acf{MiniRocket} algorithm, initially proposed by Dempster et al.~\cite{dempster21_minir}, represents an incremental improvement over the \textsc{Rocket}~\cite{dempster20_rocket} algorithm, a time-series classification method known for its speed and performance.

\ac{MiniRocket} differs from other time-series classification algorithms by applying a number of random small convolutional kernels, 84 in the original as well as in our implementation, to the input time-series, extracting one feature, and running a linear classifier. In \ac{MiniRocket}, these convolutional kernels are of length 9-time points and are randomly extracted from only two weights, \{-1,2\}. They are applied per channel on the input time series using a dot-product. From the convoluted signal, a settable number of features can be extracted. 

\ac{MiniRocket} calculates a distinct feature from each convolutional kernel: the \ac{PPV}. The features form a transformed representation of the input time series that can be input into any linear classifier or regressor.
\ac{PPV} represents the proportion of positive values in the time series after convolution with the kernel. By analyzing the proportion of positive values, statistical information can be derived about the distribution of the data, which is particularly useful for capturing patterns in the time series data.
Upon the successful generation of features, these are then fed into a linear classifier or regressor. Ridge Regression was employed by the authors of~\cite{dempster21_minir}, and was therefore chosen also in this work for its efficiency and performance. However, other linear models could be used as well.

The hyperparameter search then focused on:
\begin{description}
    \item[Sampling frequency:] The analysis was done on sampling rates ranging from \qty{10}{\hertz} to \qty{3.2}{\kilo\hertz}, the aim was to maximize F1-score.
    \item [Window dimension:] The window length was spanned from 5 to 200 samples, which yielded \qty{25}{\milli\second} to \qty{1}{\second}. The sampling frequency was fixed to the value obtained in the previous evaluation. The aim was to maximize the F1-score.
    \item [Features:] These hyperparameters have been tested for their memory impact and energy consumption, being a hard constraint on our system. Features ranged from 84 to 336, and samples from 80 to 150.
\end{description}

\subsection{Hardware}
The sensor node, named \textsc{SmartTag} and introduced in\cite{giordano21_smart,giordano22_desig_perfor_evaluat_ultral_power}, aims for ultra-low power consumption and energy efficiency. \cref{fig:system} shows it connected to a drill during its typical usage (a), and next to a drill bit and a 50 cent euro coin as a size reference (c).
It is based on a flexible architecture, which can be adapted to various workloads and easily exchange different \acp{SoC} of the \textsc{Nordic Semiconductor nRF52} family. The measurements reported in this work refer to a \textsc{SmartTag} equipped with the \textsc{nRF52810} \ac{MCU}. The \textsc{SmartTag} provides two main sensors: a temperature and humidity sensor, and an ultra-low power accelerometer. The \textsc{IIS2DLPC} accelerometer from \textsc{ST Microelectronics} was selected for its various scales of acceleration measurement and extremely low-power figures, with a current consumption of a few \unit{\micro\ampere} for measurements in low-power mode and \qty{50}{\nano\ampere} in the lowest sleep mode. 
Finally, the \textsc{SmartTag} transmits its current state together with the associated activity and cumulative runtime using \ac{BLE} advertisements at a settable interval. 

\subsection{\acs{MiniRocket} \textsc{C} implementation}

A core contribution of this paper is investigating the use of the \ac{MiniRocket} algorithm for inference on resource-constrained devices, taking as a starting point its proven lightweight training routine. This section delves into the implementation of the \ac{MiniRocket} algorithm on a \ac{MCU}.

Implementing the \ac{MiniRocket} algorithm on an \ac{MCU} presents several challenges, primarily due to the limited computational capabilities and memory of microcontrollers: The \textsc{nRF52810} is clocked at \qty{64}{\mega\hertz}, achieves an \textsc{EEMBC CoreMark\copyright} of 144 and is limited by \qty{192}{\kilo\byte} flash and \qty{24}{\kilo\byte} RAM. Furthermore, it does \textit{not} host a \ac{FPU} to perform floating-point computations.

We ported inference-related sections of the \ac{MiniRocket} library from \textsc{Python} to the \textsc{C} Language for embedded execution on the \ac{MCU}. In order to avoid any potential out-of-memory errors at runtime and to speed up execution, only static allocations were used in the developed \textsc{C} library. The porting mainly comprises the \texttt{transform} function which applies the \ac{MiniRocket} kernels to extract feature-rich embeddings, and the \texttt{predict} function which encompasses linear classification.

For the sake of completeness, both a floating-point and an integer-quantized implementation is created within the \textsc{C} library. Both implementations support binary and multi-class classification scenarios. Furthermore, independent of the number of output classes, the library also supports inference on both uni-variate and multi-variate inputs.

For the quantized implementation, 32-bit integer arithmetic is used, since it allows to maintain the same accuracy with respect to the 32-bit floating point implementation and is natively supported by the \ac{MCU}. To allow for this quantization, the inputs and the pre-trained model parameters are scaled and rounded to integers in a way that prevents value overflows during classification. More specifically, a two-step calibration procedure is performed that first addresses the quantization of the \ac{MiniRocket}'s \texttt{transform} function and subsequently of the linear classification that follows it. In the following sections, we first describe these steps and then briefly discuss the validation of the integer-quantized implementation.

\renewcommand{\arraystretch}{1.5}
\begin{table*}[t]
\centering
\begin{tabular}{cc}
\textbf{Param.}   & \textbf{Value}  \\
\hline
\hline
Samp. Rate    & \qty{200}{\hertz}    \\
\hline
Window      & \qty{80}{} samp.    \\
\hline
Kernel   & \qty{84}{}   \\
\hline
Features  & \qty{84}{} \\
\end{tabular}
\quad\quad
\begin{tabular}{cc}
\textbf{Metric}   & \textbf{Value}  \\
\hline
\hline
Flash    & \qty{7}{\kilo\byte}    \\
\hline
RAM      & \qty{3}{\kilo\byte}    \\
\hline
Latency  & \qty{8.6}{\milli\second} \\
\hline
F1-score & 0.969 
\end{tabular}
\quad\quad
\begin{tabular}{cc}
\textbf{Energy}   & \textbf{Value}  \\
\hline
\hline
Idle - \qty{1}{\second}    & \qty{4.7}{\micro\joule}    \\
\hline
Acc. samp.      & \qty{630}{\micro\joule}    \\
\hline
MiniRocket   & \qty{72}{\micro\joule}   \\
\hline
Adv.  & \qty{67}{\micro\joule} \\
\end{tabular}
\vspace{0.3cm}
\caption{Summary tables of the \ac{MiniRocket} algorithm implementation. Left: parameters selected from the hyperparameter study. Center: implementation-specific metrics of \ac{MiniRocket}. Right: energy profile of the application.}
\label{tab:summary}
\vspace{-0.8cm}
\end{table*}

\subsubsection{Quantizing the \texttt{transform} function}

Given input accelerometer readings with the largest absolute magnitude of \(I_m\) milli-G, the first calibration step involves finding the integer multiplicative factor to be used for scaling the input readings and the transform function biases. We quantize only biases since all the other pre-trained parameters used in the transform function are already of integer type. The optimal integer scale factor value is estimated so as to maximize the use of the available bit-depth while disallowing value overflows during inference. To explain how this is done, we first make the observation that, at the atomic level, the \texttt{transform} function is comprised of many comparisons of the mathematical form as in \cref{eq_comparison}.

\begin{equation}
\label{eq_comparison}
\left(\sum_{i=1}^{N_j} I_i \right) < B_j,
\end{equation}

where \(I_i\) are the input values. \(N_j\) and \(B_j\) are respectively the number of input features summed and the corresponding learned bias value used for the \(j
\)-th comparison. Note that this mathematical form holds not only for uni-variate inputs but also multi-variate ones, as the above sum can be viewed as the flattened version of two summations, one of which is over the input feature channels dimension for multi-variate inputs. Note also that given \(I_m\) and the pre-trained bias values, we can find the highest possible values that can be obtained on both sides of the comparisons by considering the largest absolute values for \(B_j\) and \(N_j\) over all the existing comparisons for a pre-trained \ac{MiniRocket} model. More formally, given a model with the maximum absolute bias value of \(B_m\) and the maximum number of input features \(N_m\), we estimate the integer scale factor with \cref{eq_scale_1}.

\begin{equation}
\label{eq_scale_1}
S_1 = \left\lfloor (2^{(b-1)} - 1) / \max(I_m \cdot N_m,\ B_m) \right\rfloor
\end{equation}

where \(\lfloor x \rfloor\) is the floor function that gives the greatest integer less than or equal to \(x\), and \(b\) is the number of bits used for integer arithmetics. In this work, we use \(b = 32\) bits and set \(I_m\) to \(16'000\) milli-G. 

Given a \ac{MiniRocket} model trained on our dataset, we quantize the bias values by multiplying them with the scale factor \(S_1\) and rounding them to the nearest integer. At inference time on the \ac{MCU}, to prevent overflows due to incorrect conditioning of the sensor readings, we first clamp the input values into the range \([-I_m,\ +I_m]\) and then multiply them with the scale factor before running the fixed-point transform function using the quantized biases.

\subsubsection{Quantizing the \texttt{predict} function}

The output of the quantized \texttt{transform} function is a feature vector \(\mathbf{t}_{1\times T}\) with a value range \([0,\ C_k]\), where \(C_k\) is the number of comparisons used for the \(k\)-th element of the feature vector. They are obtained as the sum of the \(C_k\) comparisons (i.e., each resulting in 0 or 1) of the form \cref{eq_comparison}. In the floating-point implementation, however, the feature values are normalized within the range \([0,\ 1]\) by dividing each by its respective \(C_k\) value.

As in the quantization of the \texttt{transform} function, the second calibration step involves finding the scale factor that results in the most efficient use of the available bit-depth without any value overflows. Given a pre-trained linear classifier with a weight matrix \(\mathbf{W}_{T\times K}\) and a bias vector \(\mathbf{b}_{1\times K}\), this is done by first splitting the classifier weight values into mutually exclusive positive and negative sets for each distinct output class \(k \in [1,\ K]\), and then summing the weights inside each set. Let the positive and the negative sum for the output class \(k\) be \(w^p_k\) and \(w^n_k\) respectively. The final scale factor can be obtained by \cref{eq_scale_2},

\begin{equation}
\label{eq_scale_2}
S_2 = \left\lfloor \sqrt{ (2^{(b-1)} - 1) / \max(f(k)\vert_{k = 1,\dots, K}) } \right\rfloor
\end{equation}

where \(f(k) = \max(w^p_k + \vert b_k\vert, -w^n_k + \vert b_k\vert)\) and \(b_k\) is the \(k\)-th value in the bias vector. We use \(S_2\) to quantize the input features \(\mathbf{t}_{1\times T}\) from the transform function, as well as the weight matrix \(\mathbf{W}_{T\times K}\) and the bias vector \(\mathbf{b}_{1\times K}\) of the linear classifier. Their quantization is performed using the equations \cref{eq_quant_predict}:

\begin{align}
\begin{split}
\label{eq_quant_predict}
\mathbf{t}_{1\times T}^q &= (S_2 \cdot \mathbf{t}_{1\times T} + C_k \backslash 2) \backslash C_k, \\
\mathbf{W}_{T\times K}^q &= \left\lfloor (S_2 \cdot \mathbf{W}_{T\times K}) \right\rceil, \\
\mathbf{b}_{1\times K}^q &= \left\lfloor (S_2 \cdot S_2 \cdot \mathbf{b}_{1\times K}) \right\rceil,
\end{split}
\end{align}

where \(\lfloor\cdot\rceil\) is the rounding function to the nearest integer and \(\backslash\) is the integer division operator. The output of the quantized \texttt{predict} function is an integer vector of size \(1\times K\) holding per-class scores, from which we find the highest score and return its class index \(k^* \in [1,\ K]\).

\subsubsection{Validation}

To validate the floating-point and quantized implementations, we developed a \textsc{Python} application, which trains the algorithm, runs the inference on the held-out test set, and exports the quantized kernels as well as the classifier parameters in static \textsc{C} arrays. Moreover, the application generates sample input and output data files to validate the quantized implementation on a computer or directly on the \ac{MCU}.

Tests performed on our dataset demonstrated that the floating-point implementation of our \textsc{C} library generates bit-wise equivalent results as the Python implementation for both the \texttt{transform} and \texttt{predict} function outputs. Furthermore, randomized repetitions involving re-trainings showed that the difference between the integer-quantized and floating-point implementations remained to be below \(1\%\) without any perceivable bias for all the samples in our dataset in terms of both classification accuracy and F1-score.

\subsection{\acs{MCU} runtime}
\begin{figure}[!b]
  \centering
  \includegraphics[width=0.48\textwidth]{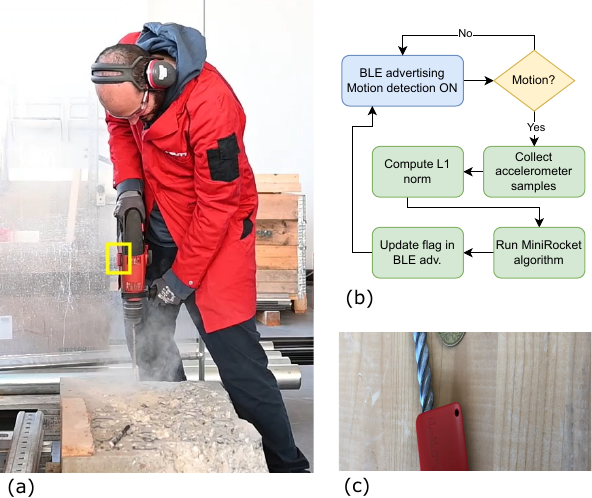}
  \caption{(a) Tool during operation with the \textsc{SmartTag} applied on it, highlighted in yellow. (b) Working principle of the \textsc{SmartTag}. (c) Close-up of the \textsc{SmartTag} device.}
  \label{fig:system}
\end{figure}

The flowchart of \textsc{SmartTag}'s operation is shown in \cref{fig:system}b). A fundamental source of power consumption in the final application is the baseline power consumption of the \ac{MCU}, \ac{BLE}, and accelerometer. In order to achieve the lowest power consumption, the algorithm was heavily duty-cycled on the \textsc{SmartTag}. The system is constantly sleeping and is woken up by external interrupts and by a timer to transmit periodically \ac{BLE} advertisement packets, in our case every \qty{7}{\second}. In particular, the external interrupt is connected to the interrupt pin of the accelerometer, which implements an internal threshold-based system to recognize movement. This feature as well as the \ac{MiniRocket} hyperparameter optimization described previously, is targeted to save power. In particular, the \ac{MiniRocket} algorithm is run only when actually needed, i.e. when the system is moved to distinguish between transport and tool usage. 

Once the \textsc{SmartTag} is woken up from the accelerometer interrupt, indicating that the tool has been moved, the system collects the accelerometer data needed to run the \ac{MiniRocket}. The \(L^1\)-norm is then computed from the three axes of the accelerometer and is then fed into the algorithm input buffer. After the buffer is full, the algorithm is run, and the result is stored in a variable. To save power, we avoid connecting the \textsc{SmartTag} to a \ac{BLE} gateway; instead, we insert the inference information into the advertisement packets, which would be sent anyway at the set interval.

As a final remark on the \textsc{SmartTag} operation, the inference is run at most once per BLE advertising window. This brings a limit in runtime estimation accuracy, making the granularity of the estimation equal to the \ac{BLE} advertisement period, \qty{7}{\second} in our case. However, in most use cases, this is not an invalidating limitation, and more granularity and \ac{BLE} responsiveness can be of course traded off with a lower battery lifetime.

\begin{figure*}[!t]
    \centering
    \begin{overpic}[width=0.4\textwidth]{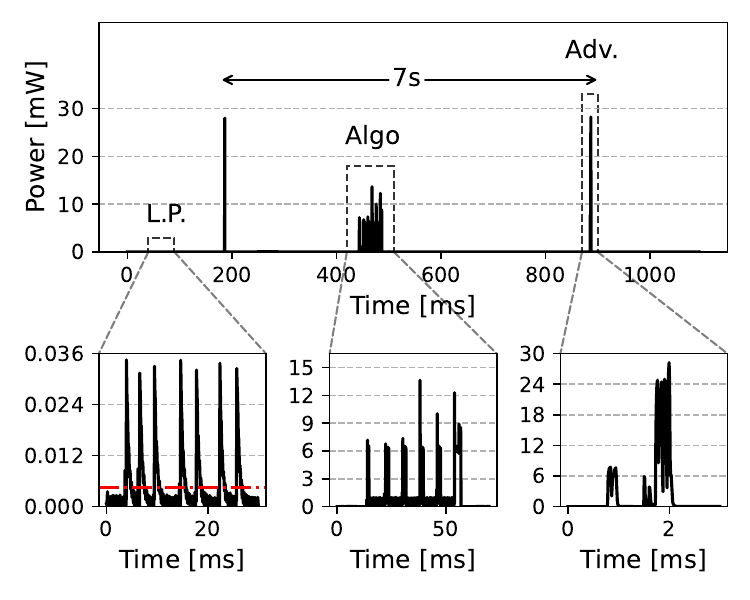}
        \put(5,1){(a)}
    \end{overpic}
    \begin{overpic}[width=0.45\textwidth]{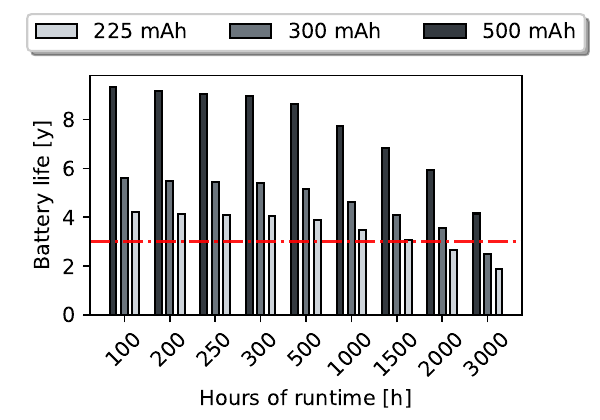}
        \put(5,1){(b)}
    \end{overpic}
  \caption{Plot (a) shows the current consumption. For the zoomed subplots, from left to right: idle, algorithm, advertising. Plot (b) shows the battery estimation for three different battery sizes and different hours of runtime.}
  \label{fig:current-battery}
\end{figure*}

\section{Results and Discussion}\label{sec:results}

In this section, we will discuss the results of the embedded implementation of the \ac{MiniRocket} algorithm in terms of classification performance, inference latency, memory, and energy consumption on the selected task, benchmarked on the target \ac{MCU}. A summary of the implementation details and the major results of this work is presented in \cref{tab:summary}.

To fulfill the goal of optimizing the algorithm in terms of memory footprint and computation overhead, we run a series of ablation studies to find out the Pareto optimum of complexity and algorithm performance. 

\subsection{Sampling frequency}
The results of the experiments can be seen in \cref{fig:algo_tuning}a). A plateau in F1-score can be seen at around \qty{200}{\hertz}, and therefore it was chosen as a sufficient sampling rate.

\subsection{Window length}
Having fixed the sampling rate (\qty{200}{\hertz}),  the dimension of the window is chosen next. \cref{fig:algo_tuning}b) reports the F1-scores across different window lengths. Following the goal to balance accuracy and memory consumption, \qty{0.4}{\second} (80 samples) was selected as the optimal window length, yielding an F1-score of 0.969.

\subsection{Feature size}
Dwelling now into the memory profiling, it was possible to fit the application into the \textsc{nRF52810} \ac{MCU} of the \textsc{SmartTag}. \ac{MiniRocket} occupies a total of \qty{7}{\kilo\byte} of flash out of the \qty{15}{\kilo\byte} left by the rest of the application (sensor reading, \ac{BLE} advertising, bootloader, \dots) and \qty{3}{\kilo\byte} of RAM in the 84 features and 80 time-samples configuration, including input and output buffers. \cref{fig:algo_tuning}c) visualizes the RAM memory profiling for different features and windowing configurations. These complexity-increasing settings could be used to target more demanding machine learning tasks, such as multi-class classification or regression tasks.

\subsection{Power profiling}
The measurements were performed with the settings resulting from the ablation studies, i.e. 84 features, \qty{200}{\hertz} sampling rate, and 80 samples time-window, yielding an F1-score of 0.969. Power has been profiled powering the \textsc{SmartTag} with \qty{3}{\volt} from the battery connectors.

\cref{fig:current-battery}a) provides a complete power profiling of the application. The \ac{BLE} advertisement peaks are clearly visible up to around \qty{30}{\milli\watt}, and they are spaced \qty{7}{\second} as designed. Moreover, in the detailed plot on the right, the accelerometer detects a movement: the system is woken up, the samples needed by \ac{MiniRocket} are collected and read out. Just after that, the \ac{MiniRocket} algorithm is run on the collected data, and the inference result will be stored in the advertisement packet to be sent out in the next advertisement cycle. In this way, we can provide an update every advertisement window, and we do not incur the hefty energy consumption of being always connected to \ac{BLE}. 

The horizontal red dashed line represents the average idle consumption, which averages at \qty{4.7}{\micro\watt}.

The methodology described above yields great energy savings, first by activating the acquisition process and \ac{ML} algorithm only when necessary by gating it with a very low-power thresholding implemented in hardware inside the accelerometer. Secondly, by running the inference only once per window. The detailed power consumption of the \ac{MiniRocket} algorithm is shown in the central zoomed window of \cref{fig:current-battery}a). 

\subsection{Battery estimation}

\cref{fig:current-battery}b) depicts the expected battery life of the smart tag when considering the following conditions: battery efficiency of \(80\%\); \ac{BLE} periodically advertising every \qty{7}{\second}; accelerometer constantly in motion detection mode at \qty{12.5}{\hertz}; accelerometer switched to \qty{200}{\hertz} for \qty{0.4}{\second} (80 samples); and \ac{ML} inference performed only once every \ac{BLE} interval (where motion is detected). Specifically, the bar chart reports the battery life estimation at different battery capacities and different covered tool runtimes.
As visible in the bar chart, battery life can go well beyond 3 years with the smallest \qty{225}{\milli\ampere{}\hour} battery, covering \(6\times\) more hours of runtime in 3 years (\qty{1500}{\hour} vs the expected \qty{250}{\hour}); or respectively 4 years can be reached covering \qty{250}{\hour} of runtime. With bigger batteries, even up to 9 years can be covered. The battery self-discharge has been taken into account considering \(80\%\) of the battery capacity. 
\section{Conclusion}\label{sec:conclusion}

This paper presented an optimized version of \ac{MiniRocket} on an on a resource-constrained \ac{MCU} hosted in a low-power sensor node in the context of activity detection on construction power tools. The collection of a dataset, training, and validation of the algorithm in a real-world scenario has been presented. Experimental evaluations have shown the capabilities of the algorithm to achieve an accuracy of \(96.9\%\) between \textit{Usage} and \textit{Transport} activities on power tools from accelerometer-only data.
The power consumption of the device running \ac{MiniRocket} could be limited to less than \qty{15}{\micro\watt} on average with an inference per hour and therefore enabling a lifetime of up to 3 years on a \qty{225}{\milli\ampere{}\hour} \textsc{CR2032} battery. Considering the margins presented by the memory and power profiling pictures, we can add that even more complex \ac{ML} tasks (e.g., multiclass, more sample, more features) can be tackled while meeting the memory size and battery life targets.
This work demonstrated it is possible to optimize the \ac{MiniRocket} interference algorithm on ultra-low power sensor nodes maintaining the same inference accuracy. This is promising to further spread the usage of TinyML on \ac{IoT} devices to enable new possibilities for Industry 4.0.

\bibliographystyle{IEEEtran}
\bibliography{bib/IEEEabrv, bib/references}
\end{document}